# Topological Phase Control via Dynamic Complex Pole-Zero Engineering

Alex Krasnok[1,2]

[1]*Department of Electrical Engineering, Florida International University, 33174, Miami, USA*
[2]*Knight Foundation School of Computing and Information Sciences, Florida International University, 33174, Miami, USA*

*To whom correspondence should be addressed:* akrasnok@fiu.edu

Precise optical phase control is crucial for innovations in telecommunications, optical computing, quantum information processing, and advanced sensing. However, conventional phase modulators often introduce parasitic amplitude modulation and struggle to provide a full 2π phase shift efficiently. This work introduces a novel paradigm for complete and robust phase control at constant amplitude by dynamically engineering the pole-zero constellation of resonant photonic systems within the complex frequency plane. We theoretically elucidate and validate two distinct approaches: first, by modulating the complex frequency of an excitation signal to trace an iso-amplitude contour (apollonian circle) around a static reflection zero; and second, by dynamically tuning the physical parameters of the resonator such that its reflection zero encircles a fixed-frequency monochromatic excitation, again constraining operation to an iso-amplitude trajectory. Both methods demonstrate the ability to impart a full 2π phase shift while maintaining a pre-defined, constant reflection amplitude, thereby eliminating amplitude-to-phase distortion. These results leverage the topological nature of phase accumulation around critical points (poles and zeros), a concept gaining significant traction in non-Hermitian and topological photonics.

***Introduction. –*** The capacity to precisely and efficiently manipulate the phase of optical signals is foundational to modern photonics [1–3]. Phase modulators are indispensable in coherent optical communication, enabling advanced modulation formats for enhanced spectral efficiency. Beyond telecommunications, they are critical for optical beam steering in LiDAR, constructing programmable photonic circuits for optical computing and machine learning, and controlling quantum states in integrated quantum photonics [4–7]. Recent explorations highlight the role of dynamic phase control in emerging areas like topological photonic memory and reconfigurable vector beam generation [8–11].

Despite their importance, conventional approaches to optical phase modulation often face inherent limitations. Many schemes based on refractive index perturbation (e.g., via electro-optic, thermo-optic, or carrier-dispersion effects) in resonant or non-resonant structures can introduce unwanted amplitude variations concomitant with the desired phase shift [4,12,13]. This parasitic amplitude modulation (AM-PM conversion) distort signals, introduce noise, and limit the performance of sensitive applications. Furthermore, achieving a full 2π phase shift, essential for complete phase control, can require substantial drive power, large device footprints, or operation over wide parameter ranges that may not always be practical or efficient, particularly in compact, integrated platforms [14–18].

The burgeoning fields of complex non-Hermitian physics and topological photonics have recently unveiled powerful new strategies for wave manipulation [19–22]. The response function of any linear, time-invariant system, such as a resonator's reflection or transmission coefficient, is defined by its poles and zeros in the complex frequency plane. The strategic placement and behavior of these critical points, influenced by concepts like non-Hermiticity and symmetry breaking [23], dictate the system's spectral and temporal response. Harnessing their topological properties, such as the quantized phase winding around them [24,25], offers novel device functionalities [26,27]. For instance, achieving a 2π phase gradient by ensuring a branch cut between a zero-pole pair crosses the real frequency axis is now understood as a key topological mechanism [28].

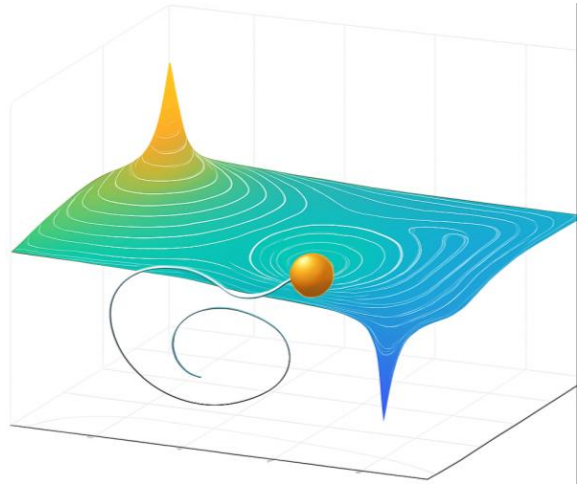

**Figure 1.** *Conceptual illustration of topological phase control in the complex frequency plane. The landscape represents a system's response function, characterized by a pole (upward peak) and a zero (downward dip). A controlled trajectory (dark spiral line) encircling the zero allows for the accumulation of a topological phase. The orange sphere indicates an operating point achieved by navigating this complex landscape, enabling precise control over the system's output phase and amplitude by exploiting the properties of these critical points.*

This work pioneers a strategy for complete (2π) phase modulation at a precisely controlled, constant amplitude by actively manipulating the interplay between an optical signal and these complex-plane critical points. The central idea, illustrated in Fig. 1, involves dynamically guiding the system's operation along specific iso-amplitude trajectories that encircle a reflection (or transmission) zero. According to Cauchy's Argument Principle [29–31], such encirclement guarantees a quantized 2π phase accumulation, a direct manifestation of the zero's topological charge. By meticulously constraining the operational path to an iso-amplitude contour, AM-PM conversion is inherently eliminated.

We investigate two distinct, yet complementary, implementations of this concept. The first approach involves maintaining a static resonant system (with fixed pole-zero locations) and dynamically shaping the complex frequency of the input optical excitation. By tailoring the real frequency and the instantaneous growth/decay rate (imaginary frequency component) of the excitation signal, it can be made to trace a pre-defined iso-amplitude circle around the system's reflection zero, thereby acquiring a 2π phase shift at constant reflected amplitude. The second, and often more practically accessible, approach utilizes a standard monochromatic input signal at a fixed real frequency. Here, the physical parameters of the resonant system itself (such as its resonant frequency and loss/gain rates) are dynamically modulated. This modulation is carefully orchestrated such that the system's reflection zero traces a path in the complex frequency plane that encircles the fixed excitation frequency, while simultaneously ensuring that the excitation frequency always lies on an iso-amplitude contour of the *instantaneously* defined system. This again results in a 2π phase shift at the desired constant reflection amplitude. Through detailed theoretical derivations and comprehensive numerical simulations, we demonstrate the efficacy of both approaches. We show that by actively engineering the pole-zero dynamics relative to the excitation, it is possible to achieve robust, complete phase control with high fidelity, paving the way for a new class of optical phase modulators with superior performance characteristics.

***Results.*** **–** The system central to our investigation is a canonical resonant photonic structure: a single optical resonator side-coupled to a waveguide. For the purpose of phase modulating the reflected optical signal, this system is effectively treated as a single-port device. An input optical wave interacts with the resonator, and the phase-modified output is the wave reflected back from the coupling point. The interaction is governed by the resonator's intrinsic properties and its coupling to the waveguide. Central to our analysis is the reflection coefficient $r(\tilde{f})$, defined in the complex frequency plane $\tilde{f} = f_R + jf_I$ (where $f_R$ is the real frequency and $f_I$ denotes the imaginary part, associated with temporal decay or amplification). For the single side-coupled resonator, the coefficient is given by (for more details, see Supplementary Materials, S1):

$$r(\tilde{f}) = \frac{j(\tilde{f} - f_0) + (\Gamma_0 - \Gamma_c)}{j(\tilde{f} - f_0) + (\Gamma_0 + \Gamma_c)} \quad (1)$$

Here, $f_0$ is the intrinsic resonant frequency of the isolated resonator, $\Gamma_0$ is the intrinsic loss rate (accounting for absorption and non-waveguide radiation losses), and $\Gamma_c$ is the external coupling rate to the waveguide. The behavior of $r(\tilde{f})$ is determined by its poles and zeros. The reflection zero $\tilde{f}_z$, satisfying $r(\tilde{f}_z) = 0$, is obtained by setting the numerator of Eq. (1) to zero, $\tilde{f}_z = f_0 + j(\Gamma_0 - \Gamma_c)$. The corresponding pole $\tilde{f}_p$, where $r(\tilde{f}_p) \to \infty$, follows from the denominator, $\tilde{f}_p = f_0 + j(\Gamma_0 + \Gamma_c)$.

All simulations assume $f_0$ at a telecom wavelength of 1.55 μm (approximately 193.41 THz). Parameters were chosen to place the system in an over-coupled regime, with $\Gamma_0 = 0.2$ THz and $\Gamma_c = 0.5$ THz, ensuring that the primary reflection zero resides in the lower half of the complex frequency plane, a favorable condition for illustrating zero-encirclement principles in phase modulation. For the parameters selected, $\tilde{f}_z \approx 193.41 - j0.3$ THz and $\tilde{f}_p \approx 193.41 + j0.7$ THz.

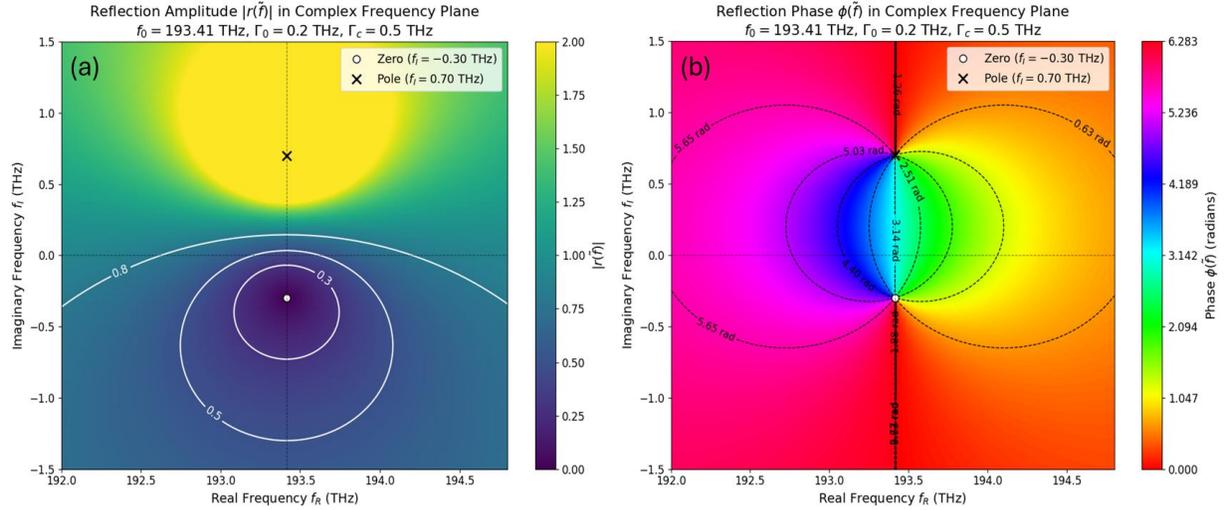

**Figure 2.** (a) Reflection amplitude $|r(\tilde{f})|$ in the complex frequency plane for a side-coupled resonator with $f_0 \approx 193.41$ THz, $\Gamma_0 = 0.2$ THz, and $\Gamma_c = 0.5$ THz. The color map indicates the amplitude, with the reflection zero (at $f_i = -0.3$ THz) and pole (at $f_i = +0.7$ THz) marked. White lines represent iso-amplitude contours for $|r| = 0.3, 0.5,$ and $0.8$. (b) Reflection phase $\phi(\tilde{f})$ (from 0 to $2\pi$ radians) in the complex frequency plane for the same resonator parameters. The color map illustrates phase, showing characteristic $2\pi$ winding around the reflection zero and pole. Dashed black lines are iso-phase contours.

These complex frequencies act as topological charges [32], shaping the system's phase response. The phase $\phi(\tilde{f})$ of the reflection coefficient is given by $\arg(\tilde{f} - \tilde{f}_z) - \arg(\tilde{f} - \tilde{f}_p)$, after incorporating the $j$ factor into the argument evaluation or redefining the pole/zero locations accordingly. According to Cauchy's Argument Principle [29–31], the total phase shift $\Delta\phi$ around a closed contour $C$ in the complex frequency plane depends on the number of enclosed zeros ($N_0$) and poles ($N_p$) (for more details, see Supplementary Materials, S2):

$$\Delta\phi_C = 2\pi(N_0 - N_p) \quad (2)$$

This principle underpins the topological robustness of the phase shifts we exploit.

Equally important are iso-amplitude contours: paths in the complex frequency plane where the reflection amplitude $|r(\tilde{f})| = C$ is constant. For Eq. (1), these form *Apollonian circles*, described by (for more details, see Supplementary Materials, S3):

$$(\Re(\tilde{f}) - f_0)^2 + \left(\Im(\tilde{f}) - \left(\Gamma_0 - \Gamma_c\frac{1+C^2}{1-C^2}\right)\right)^2 = \left(\frac{2C\Gamma_c}{|1-C^2|}\right)^2 \quad (3)$$

This circle is centered at $f_0 + j\left[\Gamma_0 - \Gamma_c\frac{1+C^2}{1-C^2}\right]$ with radius determined by $C$ and $\Gamma_c$.

Figure 2 visualizes these characteristics. Figure 2a shows a density plot of $|r(\tilde{f})|$, highlighting the zero (dark region at $f_I \approx -0.3$ THz) and the pole (bright region at $f_I \approx +0.7$ THz, amplitude capped at 2.0). Superimposed are iso-amplitude contours for $|r| = 0.3, 0.5, 0.8$, confirming their predicted circular geometry. Figure 2b presents the reflection phase $\phi(\tilde{f})$, spanning 0 to $2\pi$ radians, illustrating phase winding: $+2\pi$ around the zero, $-2\pi$ around the pole, as seen in the color gradients and dashed iso-phase lines. This confirms the viability of circular iso-amplitude paths for robust $2\pi$ phase modulation.

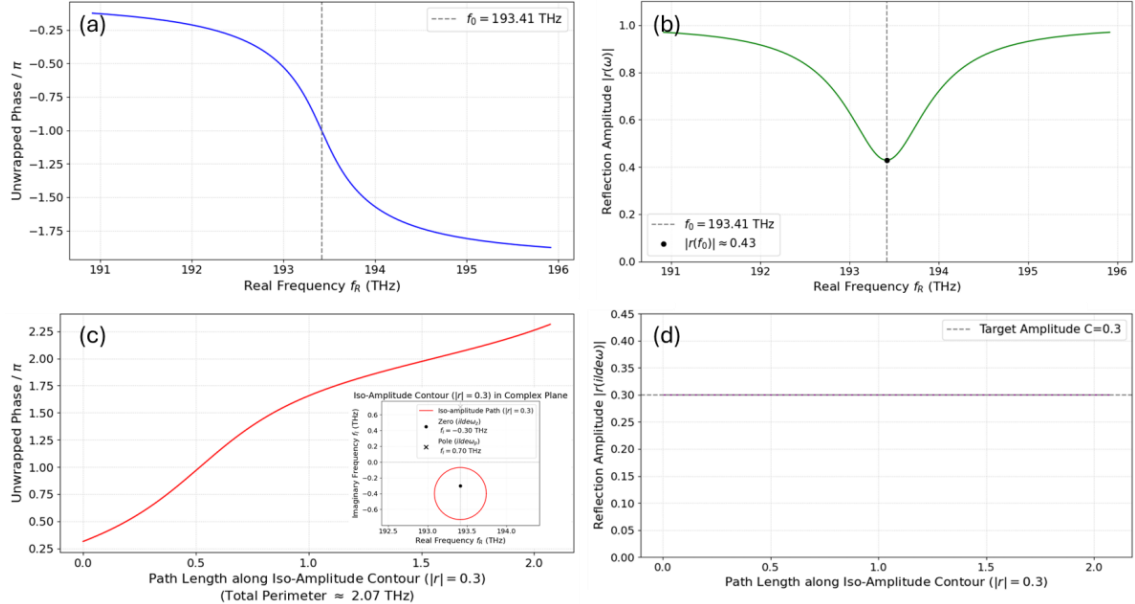

**Figure 3.** (a) Unwrapped reflection phase (normalized by $\pi$) versus real excitation frequency $f_R$ for a 5.0 THz sweep along the real axis ($f_I = 0$), centered at $f_0 \approx 193.41$ THz. The total phase change observed is approximately $-1.75\pi$ radians. (b) Reflection amplitude $|r(\omega)|$ versus real excitation frequency $f_R$ for the 5.0 THz sweep along the real axis, corresponding to the phase response in (a). The amplitude varies significantly, dipping to approximately $0.43$ at $f_0$. (c) Unwrapped reflection phase (normalized by $\pi$) versus path length along an iso-amplitude contour ($|r| = 0.3$) encircling the reflection zero (Approach 1). A complete $2.00\pi$ radian phase shift is achieved over the contour perimeter of approximately $2.07$ THz. Inset: Path of the complex excitation frequency $\tilde{f}_{exc}(t)$ in the complex frequency plane for Approach 1. The red circle represents the iso-amplitude contour ($|r| = 0.3$) which encloses the static reflection zero ($\tilde{\omega}_z$) but not the pole ($\tilde{\omega}_p$). (d) Reflection amplitude $|r(\tilde{\omega})|$ versus path length along the iso-amplitude contour ($|r| = 0.3$) encircling the reflection zero. The amplitude remains constant at the target value of $0.3$.

In the approach 1, the resonator parameters $f_0$, $\Gamma_0$, and $\Gamma_c$ are static. Phase modulation is achieved by dynamically varying the *complex excitation frequency* $\tilde{f}_{exc}(t)$. First, we contrast this with conventional real-frequency modulation. Figure 3a shows the unwrapped phase (normalized by $\pi$) of $r(\tilde{f})$ for a real-axis sweep across 5.0 THz around $f_0$. A phase excursion of about $-1.5\pi$ is observed, that is typical of resonance probing along the real axis. Yet this also induces significant amplitude variations. Figure 3b shows $|r(f_R)|$ over the sweep: it drops from near-unity far from resonance to about 0.43 at $f_0$, matching the theoretical minimum for over-coupling, $|r(f_0)| = \left|\frac{\Gamma_0 - \Gamma_c}{\Gamma_0 + \Gamma_c}\right| \approx 0.429$. Such amplitude variation exemplifies the AM-PM coupling inherent to real-axis tuning.

To address this, we instead steer $\tilde{f}_{\text{exc}}(t)$ along a specific iso-amplitude contour, here, for $|r| = 0.3$. Figure 3c (inset) explicitly plots this contour: a circle with ~2.07 THz perimeter enclosing the zero but excluding the pole. Figure 3c shows the unwrapped phase (normalized by $\pi$) during this traversal: a clean $2.00\pi$ shift, in accordance with Cauchy's Principle. Simultaneously, Figure 3d shows that $|r(\tilde{f}_{\text{exc}})| = 0.3$ remains constant, demonstrating pure phase modulation, devoid of amplitude artifacts. Compared to the 5.0 THz real-axis sweep, this method delivers a full $2\pi$ phase shift with amplitude stability and *reduced spectral path* length (~2.07 THz).

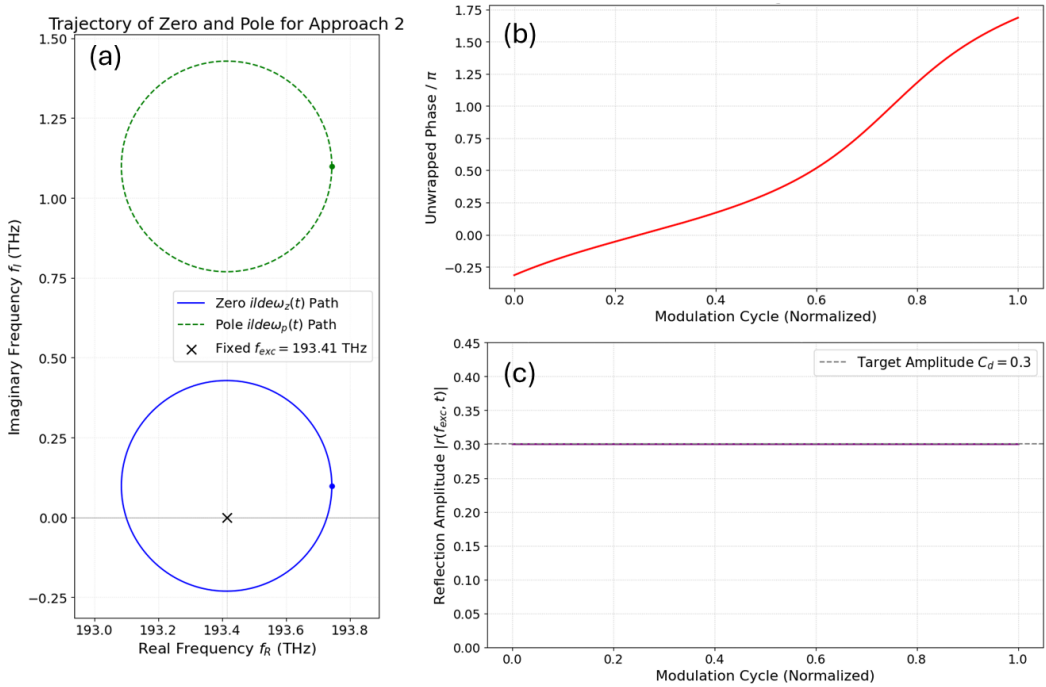

***Figure 4.*** *(a) Trajectories of the dynamically modulated reflection zero $\tilde{\omega}_z(t)$ (blue solid circle) and pole $\tilde{\omega}_p(t)$ (green dashed circle) in the complex frequency plane for Approach 2. The fixed real excitation frequency $f_{exc}$ (black 'x') is encircled by the zero's path but not by the pole's path. (b) Unwrapped reflection phase (normalized by $\pi$) at the fixed real excitation frequency ($f_{exc} \approx 193.41$ THz) during one cycle of system parameter modulation (Approach 2). A complete $2.00\pi$ radian phase shift is achieved. (c) Reflection amplitude $|r(f_{exc}, t)|$ at a fixed real excitation frequency ($f_{exc} \approx 193.41$ THz) as system parameters ($f_0(t), \Gamma_0(t)$) are modulated over one cycle (Approach 2). The amplitude remains constant at the target value $C_d = 0.3$.*

We now consider Approach 2, where the excitation frequency is fixed and the system parameters are dynamically modulated. This alternative approach uses a fixed real excitation frequency $f_{\text{exc}} = f_0 \approx 193.41$ THz and achieves phase modulation by dynamically tuning system parameters. Specifically, we vary $f_0(t)$ and $\Gamma_0(t)$, keeping $\Gamma_c = 0.5$ THz fixed, such that the reflection amplitude remains $C_d = 0.3$. The reflection coefficient becomes $r(f_{exc}, t) = \frac{f_{exc} - \tilde{f}_z(t)}{f_{exc} - \tilde{f}_p(t)}$, with $\tilde{f}_z(t) = f_0(t) + j(\Gamma_0(t) - \Gamma_c)$. The iso-amplitude constraint $|r(f_{exc}, t)| = C_d$ imposes a relation between $f_0(t)$ and $\Gamma_0(t)$, $(f_{exc} - f_0(t))^2 + (\Gamma_c - \Gamma_0(t))^2 = C_d^2[(f_{exc} - f_0(t))^2 + (\Gamma_0(t) + \Gamma_c)^2]$. This implies that $f_0(t) - f_{exc}$ and $\Gamma_0(t) - \Gamma_{0,center}$ must trace a circle in parameter space. The center is (for more details, see Supplementary Materials, S4):

$$\Gamma_{0,center} = \frac{1 + C_d^2}{1 - C_d^2}\Gamma_c \quad (4)$$

and the radius:

$$R_{param} = \frac{2C_d}{1 - C_d^2}\Gamma_c \quad (5)$$

Assuming $C_d < 1$, modulating along this circle maintains constant amplitude. Figure 4a shows the trajectory of $\tilde{f}_z(t)$ in the complex plane (solid blue line), forming a circle of radius $R_{param} \approx 0.33$ THz around $f_{exc}$ (black cross). The corresponding pole trajectory (green dashed) is offset upward by $2\Gamma_c$, thus not encircling $f_{exc}$. Figure 4b reveals a smooth $2.00\pi$ phase evolution. This is attributed to $\phi(f_{exc}, t) = \arg(f_{exc} - \tilde{f}_z(t)) - \arg(f_{exc} - \tilde{f}_p(t))$ with the $2\pi$ shift stemming mainly from $\tilde{f}_z(t)$'s encirclement of $f_{exc}$. Figure 4c confirms amplitude constancy: $|r(f_{exc}, t)| = 0.3$ throughout the cycle.

***Discussion.*** **–** The results presented demonstrate the viability of achieving complete $2\pi$ phase modulation at a precisely controlled, constant reflection amplitude by dynamically engineering the interaction between an optical signal and the reflection zeros of a resonant system. Both investigated approaches: modulating the complex frequency of the excitation signal around a static zero (Approach 1) and dynamically steering the system's zero around a fixed-frequency monochromatic excitation (Approach 2) successfully yield the desired outcome, underscoring the power of leveraging the topological properties inherent in the complex frequency plane.

A key advantage highlighted by these findings is the intrinsic decoupling of phase and amplitude modulation. In conventional phase shifters, particularly those relying on resonant enhancement, altering the phase often leads to undesirable variations in signal amplitude (AM-PM conversion), as illustrated by our real-axis frequency sweep simulations (Figure 3a,b). This parasitic amplitude modulation can severely degrade the performance of phase-sensitive applications, such as higher-order quadrature amplitude modulation (QAM) in coherent communications or interferometric measurements in sensing. The strategies presented here, by constraining operation to an iso-amplitude contour (Approach 1, Figure 3) or by dynamically co-modulating system parameters to maintain a constant reflection amplitude at the operating frequency (Approach 2, Figure 4), inherently suppress this AM-PM conversion. This leads to a higher fidelity phase modulation, which is critical for advanced optical systems.

The $2\pi$ phase accumulation observed upon encircling a zero is a direct consequence of Cauchy's Argument Principle and is topologically protected. This means that once the encirclement is established, the total phase shift is robust against small perturbations in the system parameters or the exact path taken, as long as the zero remains enclosed and no other critical points (like poles) are additionally encircled or crossed [33,34]. This topological robustness offers a significant advantage over conventional phase modulators where the accumulated phase is often highly sensitive to precise parameter values and environmental fluctuations.

Comparing the two approaches, Approach 1 (modulating the complex excitation frequency) offers a conceptually direct translation of the complex-plane analysis. The resonator remains static, simplifying device design. While the practical implementation of generating arbitrary optical waveforms with time-varying real and imaginary frequency components was historically considered a considerable technological challenge, recent advancements demonstrate that such tailored excitations are becoming an increasingly established and powerful practice in integrated photonics. For instance, Hinney et al. (2024) have recently shown that precisely shaping optical excitation signals in time, effectively creating excitations oscillating at complex frequencies, can achieve virtual critical coupling and significantly enhance energy transfer to resonators, even those not physically meeting the critical coupling condition [35]. This underscores the growing feasibility and utility of employing complex frequency excitations for advanced control of light-matter interactions. The speed of modulation in our Approach 1 would thus be increasingly determined by the capabilities of sophisticated optical arbitrary waveform generators, which continue to evolve in bandwidth and precision [36–38].

Approach 2 (modulating system parameters) utilizes a standard monochromatic input, which is readily available, and shifts the burden of complexity to the dynamic tuning of the resonator's physical properties ($f_0(t)$ and $\Gamma_0(t)$ in our simulations). Many integrated photonic platforms offer mechanisms for such tuning, including thermo-optic effects, electro-optic effects (via carrier dispersion or Pockels effect), and micro-electro-mechanical systems (MEMS) [39–45]. The successful co-modulation of $f_0(t)$ and $\Gamma_0(t)$ to trace the required path in their parameter space while satisfying the constant amplitude constraint is non-trivial and would require sophisticated control algorithms and potentially multi-parameter actuators. The modulation speed in Approach 2 would be dictated by the response time of the chosen physical tuning mechanism (e.g., electro-optic effects can be very fast, while thermo-optic effects are typically slower) [1,46]. The necessity to dynamically control intrinsic loss ($\Gamma_0(t)$) might be particularly challenging, although techniques based on controlled free-carrier absorption or coupling to auxiliary lossy elements could be envisioned [47–49]. Our simulations show that a specific circular trajectory in the $(f_0(t), \Gamma_0(t))$ parameter space can indeed achieve the desired zero encirclement at constant amplitude.

The efficiency of phase accumulation is another noteworthy aspect. In Approach 1, a full $2\pi$ phase shift was achieved over an iso-amplitude contour with a perimeter of approximately 2.07 THz (Figure 3c). In Approach 2, the dynamic modulation of system parameters caused the zero to trace a path of similar perimeter (~2.01 THz, derived from $R_{param} \approx 0.32$ THz) around the fixed excitation frequency to achieve the same $2\pi$ phase shift. In contrast, a real-axis frequency sweep of 5.0 THz yielded only a $-1.5\pi$ phase shift (Figure 3a). This suggests that leveraging the steep phase gradients around a zero through encirclement can lead to more compact parameter variations for a given phase shift compared to conventional detuning methods.

While our theoretical analysis and numerical results have been centered on the reflection coefficient $r(\tilde{f})$ of a side-coupled resonator operating in a single-port (all-pass filter) configuration, the core mathematical formalism and the principles underlying topological phase control extend beyond this specific case. In particular, a similar resonator coupled to a waveguide can be configured as a two-port notch filter, where the transmitted signal becomes the primary observable. Under suitable definitions and coupling regimes, the transmission coefficient $t(\tilde{f})$ in such a setup can adopt the same functional form as the reflection coefficient $r(\tilde{f})$ considered in our study. As a result, the strategy of tracing iso-amplitude contours around transmission zeros to achieve full $2\pi$ phase modulation at constant transmission amplitude is equally applicable.

Moreover, while our study focused on a single side-coupled resonator, the underlying principles of pole-zero engineering and topological phase accumulation are generalizable. More complex photonic structures, such as systems of coupled resonators or photonic crystals, exhibit richer pole-zero landscapes in their transmission and reflection spectra. Dynamically navigating these landscapes could enable even more sophisticated phase responses, multi-port phase control, or the realization of other non-trivial optical functions. The ability to control the imaginary part of the frequency in Approach 1, or the intrinsic loss $\Gamma_0(t)$ in Approach 2, also hints at possibilities for integrating non-Hermitian concepts, such as exceptional points, for enhanced sensitivity or unique phase behaviors, although our current study focuses on simple zero encirclement.

Finally, the practical realization of these topological phase modulators will depend on overcoming the specific challenges associated with each approach. For Approach 1, advancements in high-speed, high-fidelity optical arbitrary waveform generation are key. For Approach 2, the development of integrated photonic devices with multiple, independently, and rapidly tunable parameters, along with sophisticated real-time feedback and control systems, will be crucial. Material platforms such as silicon photonics, lithium niobate, or III–V semiconductors, which offer various tuning mechanisms, are promising candidates.

***Conclusions –*** We have proposed and validated two distinct yet complementary approaches for achieving complete 2π optical phase modulation while maintaining a constant reflection amplitude. Both methods capitalize on the topological properties of resonant photonic systems, specifically by engineering the encirclement of a reflection zero in the complex frequency plane. The first approach involves dynamically modulating the complex frequency of an input optical signal to trace an iso-amplitude contour around a static system zero. The second approach employs a fixed-frequency monochromatic input and dynamically tunes the physical parameters of the resonator itself, causing its reflection zero to encircle the input frequency while the system's response at that frequency remains on an iso-amplitude trajectory. Our simulation results consistently demonstrate that these strategies can provide a full 2π phase shift with an inherently stable reflection amplitude, thereby circumventing the deleterious amplitude-to-phase conversion common in conventional phase modulators. This work highlights the power of complex pole-zero engineering as a design paradigm for advanced photonic components. The topological nature of the phase accumulation offers robustness, and the decoupling of phase and amplitude control addresses a critical limitation in existing technologies. While practical implementation of either approach presents distinct technological challenges: sophisticated waveform generation for the former, and multi-parameter dynamic device control for the latter, the fundamental principles laid out here offer a promising route towards a new generation of high-fidelity optical phase modulators. Such devices would have a significant impact on a wide range of applications, from enhancing data capacity in optical communications and enabling precise beam control in LiDAR, to facilitating complex operations in integrated quantum photonic circuits and optical neural networks. The exploration of these topological phase engineering concepts in more complex photonic structures holds further promise for uncovering novel optical functionalities.

## Acknowledgement

The authors acknowledge financial support from the DoE and AFOSR.

# Supplementary Materials:
# Topological Phase Control via Dynamic Complex Pole-Zero Engineering

Alex Krasnok[1,2]

[1]*Department of Electrical Engineering, Florida International University, 33174, Miami, USA*
[2]*Knight Foundation School of Computing and Information Sciences, Florida International University, Miami, USA*

*To whom correspondence should be addressed: akrasnok@fiu.edu*

## S1. Temporal Coupled Mode Theory and Reflection Coefficient Derivation

Temporal Coupled Mode Theory (TCMT) serves as a robust framework for describing the interaction between electromagnetic waves and resonant systems. In the case of a single optical resonator side-coupled to a waveguide, the system can be modeled as follows. Let $a(t)$ denote the complex amplitude of the resonant mode, normalized such that $|a(t)|^2$ corresponds to the energy stored in the resonator. The resonator features an intrinsic angular resonance frequency $\omega_0 = 2\pi f_0$ and an intrinsic decay rate $\gamma_0 = 2\pi\Gamma_0$, accounting for internal losses (e.g., absorption, radiation not coupled into the waveguide). Coupling to the waveguide introduces an external decay rate $\gamma_c = 2\pi\Gamma_c$, which governs energy exchange between the waveguide and the resonator.

Let $s_{in}(t)$ and $s_{out}(t)$ be the complex amplitudes of the input and output waves in the waveguide, respectively, normalized such that $|s_{in}(t)|^2$ and $|s_{out}(t)|^2$ represent power. The governing equations for the resonator amplitude and the input-output relationship in an all-pass filter configuration (with input and output on the same port) are:

$$\frac{d}{dt}a(t) = (j\omega_0 - \gamma_0 - \gamma_c)a(t) + \sqrt{2\gamma_c}s_{in}(t) \quad \text{(S1.1)}$$

$$s_{out}(t) = -s_{in}(t) + \sqrt{2\gamma_c}a(t) \quad \text{(S1.2)}$$

The term $\sqrt{2\gamma_c}$ serves as the coupling coefficient between the waveguide and resonator modes, chosen to conserve energy in the lossless case. The $-1$ in Equation (S1.2) captures the phase shift due to direct reflection, such as from an uncoupled interface or ideal coupler, when the resonator is far from resonance.

Assuming a time-harmonic input $s_{in}(t) = S_{in}e^{j\omega t}$, the resonator responds with $a(t) = Ae^{j\omega t}$, where $\omega = 2\pi f$ is the excitation frequency. Substituting into Equation (S1.1):

$$j\omega A = (j\omega_0 - \gamma_0 - \gamma_c)A + \sqrt{2\gamma_c}S_{in} \quad (S1.3)$$

Solving for $A$:

$$A(j\omega - j\omega_0 + \gamma_0 + \gamma_c) = \sqrt{2\gamma_c}S_{in} \Rightarrow A = \frac{\sqrt{2\gamma_c}}{j(\omega - \omega_0) + (\gamma_0 + \gamma_c)}S_{in}$$

The reflection coefficient $r(\omega) = \frac{s_{out}}{s_{in}}$ follows from substituting $A$ into Equation (S1.2):

$$r(\omega) = -1 + \sqrt{2\gamma_c}\frac{\sqrt{2\gamma_c}}{j(\omega - \omega_0) + (\gamma_0 + \gamma_c)} \quad (S1.4)$$

Combining terms over a common denominator:

$$r(\omega) = \frac{-j(\omega - \omega_0) - \gamma_0 - \gamma_c + 2\gamma_c}{j(\omega - \omega_0) + (\gamma_0 + \gamma_c)} = \frac{-j(\omega - \omega_0) - \gamma_0 + \gamma_c}{j(\omega - \omega_0) + (\gamma_0 + \gamma_c)}$$

Multiplying numerator and denominator by $j/j = 1$:

$$r(\omega) = \frac{j(-j(\omega - \omega_0) - \gamma_0 + \gamma_c)}{j(j(\omega - \omega_0) + (\gamma_c + \gamma_0))} \quad (S1.5)$$

This expression is uncommon. To match the main text, factor out $j$ from both numerator and denominator in Equation (S1.5):

$$r(\omega) = \frac{j(\omega - \omega_0) + (\gamma_0 - \gamma_c)}{j(\omega - \omega_0) + (\gamma_0 + \gamma_c)} \quad (S1.6)$$

This corresponds to Equation (1) in the main text, where $f$ replaces $\omega$, and $\Gamma$ replaces $\gamma$:

$$r(f) = \frac{j(f - f_0) + (\Gamma_0 - \Gamma_c)}{j(f - f_0) + (\Gamma_0 + \Gamma_c)} \quad (S1.7)$$

This is the system's reflection coefficient. Its poles and zeros determine the response behavior.

The zero $\tilde{f}_z$ occurs when the numerator vanishes:

$$j(\tilde{f}_z - f_0) + (\Gamma_0 - \Gamma_c) = 0 \Rightarrow \tilde{f}_z = f_0 + j(\Gamma_0 - \Gamma_c)$$

The pole $\tilde{f}_p$ is found when the denominator vanishes:

$$j(\tilde{f}_p - f_0) + (\Gamma_0 + \Gamma_c) = 0 \Rightarrow \tilde{f}_p = f_0 + j(\Gamma_0 + \Gamma_c)$$

## S2. Cauchy's Argument Principle

Cauchy's Argument Principle is a foundational result from complex analysis that links the number of zeros and poles of a meromorphic function enclosed by a closed contour to the total phase change of the function along that contour. Let $F(\tilde{f})$ be a meromorphic function, that is, analytic except for isolated poles, defined inside and on a simple closed contour $C$ in the complex frequency plane $\tilde{f}$. Suppose $F(\tilde{f})$ has no zeros or poles exactly on the contour. Let $N_0$ denote the number of zeros and $N_p$ the number of poles of $F(\tilde{f})$ inside $C$, each counted with multiplicity or order. Then, Cauchy's Argument Principle states:

$$\frac{1}{2\pi j} \oint_C \frac{F'(\tilde{f})}{F(\tilde{f})} d\tilde{f} = N_0 - N_p \quad \text{(S2.1)}$$

The integral $\oint_C \frac{F'(\tilde{f})}{F(\tilde{f})} d\tilde{f}$ represents the net change in $\ln(F(\tilde{f}))$ as $\tilde{f}$ traverses the contour $C$. Writing $\ln(F(\tilde{f})) = \ln|F(\tilde{f})| + j\arg(F(\tilde{f}))$, we observe that the imaginary part of this integral corresponds to the total accumulated phase (argument) change of the function:

$$\Delta_C \arg(F(\tilde{f})) = \Im\left(\oint_C \frac{F'(\tilde{f})}{F(\tilde{f})} d\tilde{f}\right)$$

From Equation (S2.1), we have:

$$\oint_C \frac{F'(\tilde{f})}{F(\tilde{f})} d\tilde{f} = 2\pi j(N_0 - N_p)$$

Taking the imaginary part of both sides:

$$\Im\left(2\pi j(N_0 - N_p)\right) = 2\pi(N_0 - N_p)$$

Hence, the total phase change of $F(\tilde{f})$ along the contour $C$ (traversed counter-clockwise) is:

$$\Delta_C \phi_F = \Delta_C \arg(F(\tilde{f})) = 2\pi(N_0 - N_p)$$

This result is essential in interpreting the topological nature of phase accumulation. Specifically, if a contour encloses a net excess of one zero over poles, the phase of $F(\tilde{f})$ increases by $2\pi$; if it encloses one more pole than zero, the phase decreases by $2\pi$.

For the reflection coefficient $r(\tilde{f})$, if the contour encloses a single zero ($N_0 = 1, N_p = 0$), then $\Delta_C \arg(r(\tilde{f})) = +2\pi$. Conversely, if it encloses only a single pole ($N_0 = 0, N_p = 1$), the phase change is $-2\pi$.

## S3. Apollonian Circles as Iso-Amplitude Contours

An iso-amplitude contour in the complex frequency plane is a locus of points where the magnitude of the reflection coefficient $|r(\tilde{f})|$ remains constant. Denote this constant value by $C$. Using the pole-zero form of the reflection coefficient, where the $j$ factor is absorbed or expressed as

$$r(\tilde{f}) = \frac{\tilde{f} - \tilde{f}_{z'}}{\tilde{f} - \tilde{f}_{p'}} \quad \text{(S3.1)}$$

with

$$\tilde{f}_{z'} = f_0 - j(\Gamma_0 - \Gamma_c), \quad \tilde{f}_{p'} = f_0 - j(\Gamma_0 + \Gamma_c),$$

the iso-amplitude condition becomes:

$$\left| \frac{\tilde{f} - \tilde{f}_{z'}}{\tilde{f} - \tilde{f}_{p'}} \right| = C$$

Squaring both sides:

$$|\tilde{f} - \tilde{f}_{z'}|^2 = C^2 |\tilde{f} - \tilde{f}_{p'}|^2 \quad \text{(S3.2)}$$

This is the defining equation for a *Circle of Apollonius*, the set of points $P$ in the complex plane for which the ratio of distances to two fixed points ($\tilde{f}_{z'}$ and $\tilde{f}_{p'}$) is constant.

Let

$$\tilde{f} = x + jy, \quad \tilde{f}_{z'} = x_z + jy_z, \quad \tilde{f}_{p'} = x_p + jy_p.$$

Then Equation (S3.2) becomes:

$$(x - x_z)^2 + (y - y_z)^2 = C^2[(x - x_p)^2 + (y - y_p)^2]$$

Expanding both sides:

$$x^2 - 2xx_z + x_z^2 + y^2 - 2yy_z + y_z^2 = C^2(x^2 - 2xx_p + x_p^2 + y^2 - 2yy_p + y_p^2)$$

Group like terms:

$$x^2(1 - C^2) - 2x(x_z - C^2 x_p) + y^2(1 - C^2) - 2y(y_z - C^2 y_p) + (x_z^2 + y_z^2 - C^2(x_p^2 + y_p^2)) = 0$$

If $C = 1$, the terms with $x^2$ and $y^2$ vanish, reducing the equation to a straight line, the perpendicular bisector between $\tilde{f}_{z'}$ and $\tilde{f}_{p'}$.

For $C \neq 1$, divide the entire expression by $(1 - C^2)$:

$$x^2 - 2x\frac{x_z - C^2 x_p}{1 - C^2} + y^2 - 2y\frac{y_z - C^2 y_p}{1 - C^2} + \frac{x_z^2 + y_z^2 - C^2(x_p^2 + y_p^2)}{1 - C^2} = 0$$

This is the equation of a circle. Completing the square in $x$ and $y$ allows identifying its center and radius.

The *center* of the Apollonian circle, $\tilde{f}_{center'}$, has coordinates:

$$\Re(\tilde{f}_{center'}) = \frac{x_z - C^2 x_p}{1 - C^2}$$

$$\Im(\tilde{f}_{center'}) = \frac{y_z - C^2 y_p}{1 - C^2}$$

The *radius* $R_C$ is given by:

$$R_C = \frac{C}{|1 - C^2|}|\tilde{f}_{z'} - \tilde{f}_{p'}|$$

Now apply this to our specific form of the reflection coefficient:

$$r(f) = \frac{j(f - f_0) + (\Gamma_0 - \Gamma_c)}{j(f - f_0) + (\Gamma_0 + \Gamma_c)}$$

Define $\Delta\tilde{f} = \tilde{f} - f_0$. Then:

$$r(\tilde{f}) = \frac{j\Delta\tilde{f} + A}{j\Delta\tilde{f} + B}, \quad \text{where } A = \Gamma_0 - \Gamma_c, \ B = \Gamma_0 + \Gamma_c$$

Let $j\Delta\tilde{f} = Z = Z_R + jZ_I = -\Im(\Delta\tilde{f}) + j\Re(\Delta\tilde{f})$. Then:

$$|r(\tilde{f})|^2 = \frac{|Z + A|^2}{|Z + B|^2} = C^2$$

This implies:

$$\frac{(-\Im(\Delta\tilde{f}) + A)^2 + (\Re(\Delta\tilde{f}))^2}{(-\Im(\Delta\tilde{f}) + B)^2 + (\Re(\Delta\tilde{f}))^2} = C^2$$

Rearranging leads to the Apollonian circle form given by Equation (3) of the main text:

$$(\Re(\tilde{f}) - f_0)^2 + \left(\Im(\tilde{f}) - \left(\Gamma_0 - \Gamma_c\frac{1 + C^2}{1 - C^2}\right)\right)^2 = \left(\frac{2C\Gamma_c}{|1 - C^2|}\right)^2$$

This confirms that iso-amplitude contours in the complex frequency plane are indeed Apollonian circles centered along the line $\Re(\tilde{f}) = f_0$.

## S4. Derivation of $\Gamma_{0,center}^{param}$ and $R_{param}$ for Approach 2

In Approach 2, we fix the real excitation frequency $f_{exc}$ and the desired reflection amplitude $C_d$. We also assume the coupling rate $\Gamma_c(t)$ is constant, $\Gamma_c(t) = \Gamma_c^{const}$. We then modulate the resonator's parameters $f_0(t)$ and $\Gamma_0(t)$ to satisfy the iso-amplitude constraint:

$$|\tilde{f}_{\text{exc}} - (f_0(t) + j(\Gamma_0(t) - \Gamma_c^{\text{const}}))|^2 = C_d^2 \cdot |f_{\text{exc}} - (f_0(t) + j(\Gamma_0(t) + \Gamma_c^{\text{const}}))|^2 \quad \text{(S4.1)}$$

$$(f_{\text{exc}} - f_0(t))^2 + (\Gamma_c^{\text{const}} - \Gamma_0(t))^2 = C_d^2[(f_{\text{exc}} - f_0(t))^2 + (\Gamma_0(t) + \Gamma_c^{\text{const}})^2]$$

Define $X(t) = f_{\text{exc}} - f_0(t)$, $Y(t) = \Gamma_0(t)$, and rewrite:

$$(-X(t))^2 + (\Gamma_c^{\text{const}} - Y(t))^2 = C_d^2[(-X(t))^2 + (Y(t) + \Gamma_c^{\text{const}})^2]$$

$$X(t)^2 + (\Gamma_c^{\text{const}})^2 - 2\Gamma_c^{\text{const}}Y(t) + Y(t)^2$$
$$= C_d^2[X(t)^2 + Y(t)^2 + 2\Gamma_c^{\text{const}}Y(t) + (\Gamma_c^{\text{const}})^2] \quad \text{(S4.2)}$$

Collecting terms:

$$X(t)^2(1 - C_d^2) + Y(t)^2(1 - C_d^2) - 2\Gamma_c^{\text{const}}Y(t)(1 + C_d^2) + (\Gamma_c^{\text{const}})^2(1 - C_d^2) = 0$$

Assuming $C_d \neq 1$, divide by $(1 - C_d^2)$:

$$X(t)^2 + Y(t)^2 - 2Y(t)\left(\Gamma_c^{\text{const}}\frac{1 + C_d^2}{1 - C_d^2}\right) + (\Gamma_c^{\text{const}})^2 = 0 \quad \text{(S4.3)}$$

$$(f_0(t) - f_{\text{exc}})^2 + \left(\Gamma_0(t) - \Gamma_c^{\text{const}}\frac{1 + C_d^2}{1 - C_d^2}\right)^2 = \left(\Gamma_c^{\text{const}}\frac{1 + C_d^2}{1 - C_d^2}\right)^2 - (\Gamma_c^{\text{const}})^2$$

This is the equation of a circle in the $(f_0(t), \Gamma_0(t))$ parameter space. The center of this circle is:

$$f_{0,\text{center}}^{\text{param}} = f_{\text{exc}}$$

$$\Gamma_{0,\text{center}}^{\text{param}} = \Gamma_c^{\text{const}} \cdot \frac{1 + C_d^2}{1 - C_d^2} \quad \text{(S4.4)}$$

The squared radius of this circle is $R_{\text{param}}^2$:

$$R_{\text{param}}^2 = \left(\Gamma_c^{\text{const}} \cdot \frac{1 + C_d^2}{1 - C_d^2}\right)^2 - (\Gamma_c^{\text{const}})^2 \quad \text{(S4.5)}$$

$$R_{\text{param}}^2 = (\Gamma_c^{\text{const}})^2\left[\left(\frac{1 + C_d^2}{1 - C_d^2}\right)^2 - 1\right]$$

$$R_{\text{param}}^2 = (\Gamma_c^{\text{const}})^2\left[\frac{(1 + C_d^2)^2 - (1 - C_d^2)^2}{(1 - C_d^2)^2}\right]$$

Note: $(1 + C_d^2)^2 - (1 - C_d^2)^2 = (1 + 2C_d^2 + C_d^4) - (1 - 2C_d^2 + C_d^4) = 4C_d^2$

So:

$$R_{\text{param}}^2 = (\Gamma_c^{\text{const}})^2 \cdot \frac{4C_d^2}{(1 - C_d^2)^2} \quad \text{(S4.6)}$$

Taking the square root (assuming $C_d < 1$, so $1 - C_d^2 > 0$):

$$R_{\text{param}} = \frac{2C_d}{1 - C_d^2} \cdot \Gamma_c^{\text{const}} \quad \text{(S4.7)}$$

Equations (S4.4) and (S4.7) are the formulas for $\Gamma_{0,\text{center}}^{\text{param}}$ and $R_{\text{param}}$ (Equations 4 and 5 in the main text) that define the circular path $f_0(t)$ and $\Gamma_0(t)$ must trace in their parameter space to ensure constant reflection amplitude $C_d$ at the fixed excitation frequency $f_{\text{exc}}$.